\documentclass[12pt,preprint]{aastex}

\usepackage{times}
\usepackage{graphicx}

\shorttitle{Fermi-LAT Observations of Vela-X}
\shortauthors{Abdo et al.}

\begin{document}

\title{\emph{Fermi} Large Area Telescope observations\\
	of the Vela-X Pulsar Wind Nebula}


\author{
A.~A.~Abdo\altaffilmark{2,3}, 
M.~Ackermann\altaffilmark{4}, 
M.~Ajello\altaffilmark{4}, 
A.~Allafort\altaffilmark{4}, 
L.~Baldini\altaffilmark{5}, 
J.~Ballet\altaffilmark{6}, 
G.~Barbiellini\altaffilmark{7,8}, 
D.~Bastieri\altaffilmark{9,10}, 
K.~Bechtol\altaffilmark{4}, 
R.~Bellazzini\altaffilmark{5}
% ...
}
% ...
\altaffiltext{2}{Space Science Division, Naval Research Laboratory, Washington, DC 20375, USA}
\altaffiltext{3}{National Research Council Research Associate, National Academy of Sciences, Washington, DC 20001, USA}
\altaffiltext{4}{W. W. Hansen Experimental Physics Laboratory, Kavli Institute for Particle Astrophysics and Cosmology, Department of Physics and SLAC National Accelerator Laboratory, Stanford University, Stanford, CA 94305, USA}
\altaffiltext{5}{Istituto Nazionale di Fisica Nucleare, Sezione di Pisa, I-56127 Pisa, Italy}
\altaffiltext{6}{Laboratoire AIM, CEA-IRFU/CNRS/Universit\'e Paris Diderot, Service d'Astrophysique, CEA Saclay, 91191 Gif sur Yvette, France}
\altaffiltext{7}{Istituto Nazionale di Fisica Nucleare, Sezione di Trieste, I-34127 Trieste, Italy}
\altaffiltext{8}{Dipartimento di Fisica, Universit\`a di Trieste, I-34127 Trieste, Italy}
\altaffiltext{9}{Istituto Nazionale di Fisica Nucleare, Sezione di Padova, I-35131 Padova, Italy}
\altaffiltext{10}{Dipartimento di Fisica ``G. Galilei", Universit\`a di Padova, I-35131 Padova, Italy}
% ...

\begin{abstract}
  We report on gamma-ray observations in the off-pulse window of the
  Vela pulsar PSR~B0833$-$45, using 11~months of survey data from the
  {\it Fermi} Large Area Telescope (LAT). This pulsar is located in
  the 8$^{\circ}$ diameter Vela supernova remnant, which contains
  several regions of non-thermal emission detected in the radio, X-ray
  and gamma-ray bands. The gamma-ray emission detected by the LAT lies
  within one of these regions, the $2^{\circ} \times 3^{\circ}$ area
  south of the pulsar known as Vela-X. The LAT flux is significantly
  spatially extended with a best-fit radius of $0.88^{\circ} \pm
  0.12^{\circ}$ for an assumed radially symmetric uniform disk. The
  200 MeV to 20~GeV LAT spectrum of this source is well described by a
  power-law with a spectral index of $2.41 \pm 0.09 \pm
  0.15$ and integral flux above 100 MeV of $(4.73 \pm
  0.63 \pm 1.32) \times 10^{-7}$ cm$^{-2}$~s$^{-1}$. The first 
errors represent the statistical error on the fit parameters, while the second ones 
are the systematic uncertainties. 
Detailed morphological and spectral analyses give strong constraints 
on the energetics and magnetic field of the pulsar wind nebula (PWN) 
system and favor a scenario with two distinct electron populations. 
\end{abstract}


\section{Introduction}

The Vela pulsar (PSR\,B0833$-$45) at a distance of 290~pc \citep{dod03} is one of
the closest pulsars to Earth and is therefore studied in great
detail. Its period of 89~ms and characteristic age of $\tau_{c}$ =
11,000 years make it an archetype of the class of 
adolescent pulsars. As with most other pulsars, 
the Vela pulsar was first detected through radio
observations~\citep{VelaPulsarDiscovery} and gamma-rays~\citep{VelaPulsarSASII}, 
but later studied in detail in the optical~\citep{VelaPulsarOptical}, X-ray~\citep{VelaPulsarXRay}
and gamma-ray bands~(\cite{VelaPulsarCOSB} and
\cite{VelaPulsarEGRET}). The pulsar has a spin-down energy loss rate
of $7 \times 10^{36}$ ergs s$^{-1}$ with the peak electromagnetic power emitted 
in the GeV gamma-ray band. Indeed, the Vela pulsar is the brightest steady 
astrophysical source for the Fermi-LAT~\citep{VelaPulsarLAT}. The gamma-ray properties 
of the pulsar have been studied in detail with the Fermi-LAT, locating
the gamma-ray emission far out in the magnetosphere close to the last
open field-lines.

Yet $\sim$99\% of the pulsar spindown luminosity is not observed as pulsed
photon emission and is apparently carried away as a magnetized particle 
wind.  Radio and X-ray observations established the presence of large scale
diffuse emission surrounding PSR\,B0833$-$45 -- thought to be related
to the Vela supernova remnant (SNR)~\citep{Dwarakanath, Duncan, Aschenbach}. 
These radio observations show that the roughly 8$^{\circ}$ diameter Vela 
SNR~\citep{Aschenbach} contains three distinct central regions 
of bright diffuse emission, dubbed Vela-X, Vela-Y and
Vela-Z~\citep{Rishbeth1958}. The most intense of these, Vela-X, is an
extremely bright ($\sim 1000$ Jy) diffuse radio structure of size
$2^{\circ}-3^{\circ}$ located close to PSR\,B0833$-$45. Its radio
spectral index is significantly harder than those of Vela-Y and
Vela-Z, pointing to a young population of non-thermal electrons.
Indeed, the flat radio spectral index, the proximity to the
Vela pulsar, and the large degree of radio polarization in Vela-X led~\cite{WeilerPanagia} 
to first suggest that the diffuse radio emission is a PWN formed by a relativistic outflow
powered by the spin-down of PSR\,B0833$-$45. The
deceleration of the pulsar-driven wind as it sweeps up ejecta from the
supernova explosion generates a termination shock at which the
particles are pitch-angle scattered and further accelerated to 
ultra-relativistic energies. The PWN emission extends
across the electromagnetic spectrum in synchrotron and
inverse Compton components from radio to TeV
energies~\citep{Gaensler2006}. PWNe studies can supply information on particle
acceleration in shocks, on evolution of the pulsar spindown and on
the ambient interstellar gas. 

\begin{figure*}[ht!!]
\begin{center}
\includegraphics[angle=0,scale=.7]{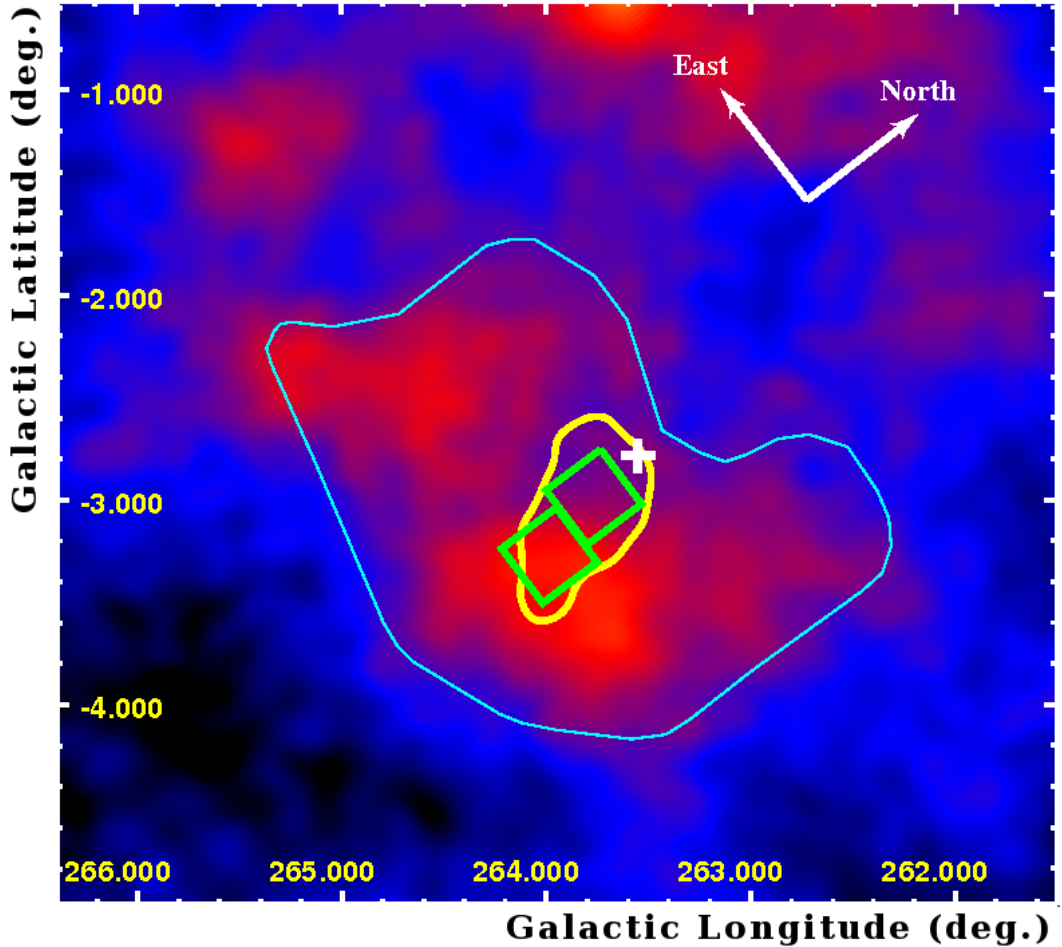} 
\caption{61-GHz WMAP (archival data) radio sky map of Vela-X in galactic coordinates. 
The position of the Vela pulsar is marked with a cross. 
The blue outer contour shows the region where the integral flux densities and the spectral indexes were computed 
for the radio data. 
The extraction regions for the spectral analysis of the ASCA data are delimited with green boxes. 
The yellow inner line presents the HESS contour at 68\% of the peak value.}
\label{fig:regions}
\end{center}
\end{figure*}

High angular resolution observations of Vela-X in different
wavebands showed a rather complex morphology. X-ray images taken 
with the Chandra X-ray telescope revealed further details~\citep{helfand}: 
two toroidal arcs of emission, 17'' and 30'' away from the pulsar, and a $4$' long collimated feature 
along the pulsar spin axis, which is interpreted as a jet. 
These structures are embedded in an extended nebula located to the south of the Vela pulsar and observed in soft X-rays 
with the ROSAT X-ray telescope. This bright X-ray and radio structure, usually referred to as the ``cocoon'', 
has an extension of $\sim 0.5^\circ \times 1.5^\circ$ and is generally 
thought to represent PWN flow crushed by the passage of the SNR reverse shock.
The offset of the cocoon to the south of the pulsar 
is explained by dense material to the north of PSR\,B0833$-$45 that 
prevents a symmetric expansion of the PWN~\citep{Blondin}. The X-ray spectrum 
of the cocoon shows a thermal component with a high-energy power-law
tail. The detection of VHE gamma-ray emission~\citep{HESSVelaX} in
the cocoon region, albeit at larger angular scales ($58' \times 43'$; yellow inner contour in Figure~\ref{fig:regions}),
clearly confirmed the notion of a non-thermal particle population in this structure. 
However, these particles do not easily explain the larger and brighter
Vela-X radio emission in the surrounding ``halo'' (blue outer contour in Figure~\ref{fig:regions}). 
This led ~\cite{dejager} to suggest a model with two 
populations of electrons: one at high energies located on the smaller cocoon 
scale, responsible for the X-ray and TeV emission and a second lower
energy population extending to larger scales and producing the
radio flux.  These models made a clear prediction that the
radio-emitting electrons should be visible in the LAT band through
inverse Compton scattering of the radio-emitting electrons off ambient
photon fields. EGRET, the predecessor of the \emph{Fermi}-LAT, was only able to place
upper limits on non-pulsed emission from this region \citep{VelaPulsarEGRET}. Recently, using the AGILE satellite, 
\cite{agile} reported the detection of the Vela pulsar wind nebula in the energy range from 100 MeV to 3 GeV.

Here we report on detection of a significant signal in the Vela pulsar 
off-pulse emission using 11 months of survey observations with the \emph{Fermi}-LAT.

\section{LAT description and observations}
\label{lat}
The LAT is a gamma-ray telescope that detects photons by conversion into 
electron-positron pairs and operates in the energy range between 20 MeV and 300 GeV. It is made of a 
high-resolution converter tracker (direction measurement of the incident 
gamma-rays), a CsI(Tl) crystal calorimeter (energy measurement) and an 
anti-coincidence detector to identify the background of charged particles 
(\cite{Atwood et al. 2009}). In comparison to EGRET, the 
LAT has a larger effective area ($\sim$ 8000 cm$^{2}$ on-axis above 1~GeV), a broader 
field of view ($\sim$ 2.4 sr) and a superior angular resolution ($\sim$ 
0.6$^{\circ}$ 68$\%$ containment at 1 GeV for events converting in the 
front section of the tracker). Details of the instruments and data processing 
are given in~\cite{Atwood et al. 2009} . The on-orbit calibration is described in 
\cite{VelaPulsarLAT}.

The following analysis was performed using 11 months of data collected
starting August 4, 2008, and extending until July 4, 2009. Only
gamma-rays in the {\emph{Diffuse}} class events were selected (with
the tightest background rejection), and from this sample, we excluded
those coming from a zenith angle larger than 105$^{\circ}$ to the
detector axis because of the possible contamination from Earth albedo
photons. We have used P6$\_$V3 post-launch instrument response
functions (IRFs), that take into account pile-up and accidental
coincidence effects in the detector subsystems\footnote{See http://fermi.gsfc.nasa.gov/ssc/data/analysis/documentation/Cicerone/Cicerone\_LAT\_IRFs/IRF\_overview.html for more details}.

\section{Timing solution}
\label{radio}
The Vela pulsar is the brightest persistent point source in the gamma-ray 
sky with pulsed photons observed up to 25~GeV. The study of Vela-X thus requires
us to assign a phase to the gamma-ray photons and select those in an off-pulse window. 
Since the Vela pulsar is young and exhibits substantial timing irregularities, 
phase assignment generally requires a contemporary radio ephemeris; such a timing model
is produced from observations made with the Parkes 64m radio telescope. However, Vela is
sufficiently bright to be timed directly in the gamma-rays; for this work and that 
reported in~\cite{velaII}, we chose to use a timing model derived directly from 
LAT observations. We used six gamma-ray times of arrival (TOA) covering 
the commissioning phase of the mission (2008 June 25 through August 4) at 5 day intervals and 
24 TOAs spaced at 2-week intervals during the survey portion of the mission 
(2008 August 4 through 2009 July 15). The TOAs were fitted to a timing model using TEMPO2; 
the RMS residuals of the TOAs with respect to the fitted model is
63~$\mu$s. More details can be found in~\cite{velaII}. Pulse phases were assigned to the LAT data 
using the {\it Fermi} plug-in provided by the LAT team 
and distributed with TEMPO2. As shown in Figure~1 of \cite{VelaPulsarLAT}
the pulsar emission is quite faint in the phase interval $\phi =0.7 -
1.0$. We have used this phase interval for both the spectral and
morphological analysis.

\section{Results}
\label{results}

The spatial and spectral analysis of the gamma-ray emission was
performed using two different methods.  The first is a
maximum-likelihood method \citep{Mattox et al. 1996} implemented in
the \emph{Fermi} SSC science tools as the ``gtlike'' code. The second
is an analysis tool developed by the LAT team called
``Sourcelike''. In the latter, likelihood fitting is iterated to the
data set to simultaneously optimise the position and the extension of
a source, assuming spatially extended source models and taking into
account nearby sources as well as Galactic diffuse and isotropic components in the fits. 
Here, we tried both point source and uniform disk models. Sourcelike can also be used to assess the 
Test Statistic (TS) value and to compute the spectra of both extended and point-like sources. In this method, 
the maximum likelihood is performed in independent energy bands, using a region of interest whose size is 
energy dependent: from $15^{\circ}$ at 200~MeV to $3.5^{\circ}$ at 50~GeV. 

We used the map cube file gll$\_$iem$\_$v02.fit to model the Galactic diffuse emission 
together with the corresponding tabulated model isotropic$\_$iem$\_$v02.txt 
for the extragalactic diffuse and the residual instrument emissions\footnote{Available from http://fermi.gsfc.nasa.gov/ssc/data/access/lat/BackgroundModels.html}. Other versions 
of the Galactic diffuse models, generated by GALPROP, are also used to assess systematic 
errors as discussed in section~\ref{spectrum}. Nearby sources in the field of view with 
a statistical significance larger than $5 \, \sigma$ are extracted 
as described in \cite{BrightSourceList} and taken into account in the study.

\subsection{Morphology}
\label{morpho}
\begin{figure*}
\begin{center}
\includegraphics[scale=0.7]{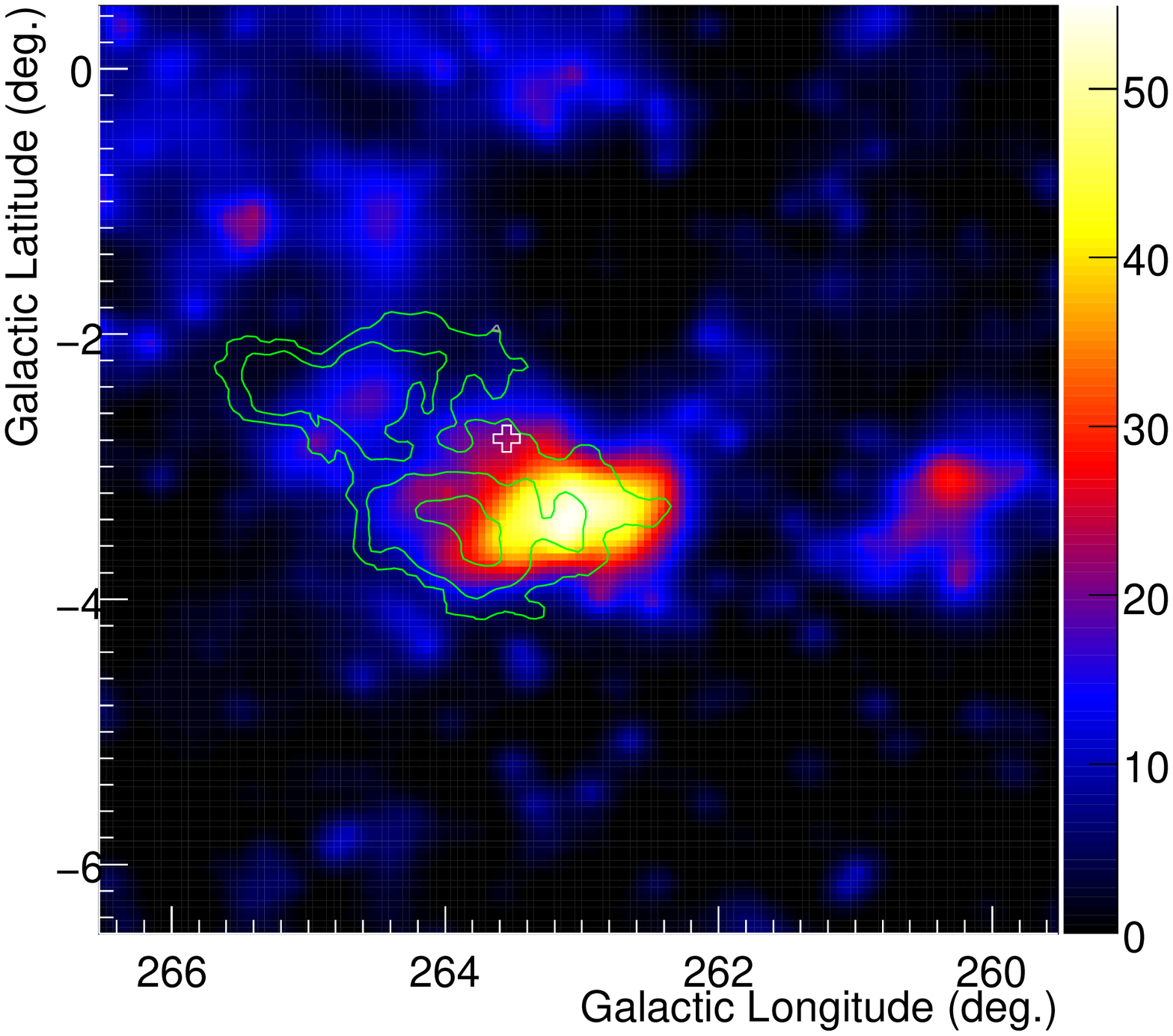}
\caption{Test Statistic (TS) map of the PWN Vela-X with side-length $7^{\circ}$, 
above 800~MeV. Each pixel of this image contains the TS value for the assumption 
of a point-source located at the pixel position. WMAP radio contours at 61~GHz 
(archival data, see section~\ref{mw}) are overlayed as green solid lines. 
The position of the Vela pulsar, PSR B0833$-$45, is indicated with a white cross. 
An unidentified source, coincident with the North-Eastern part of Puppis~A is visible 
at position (l, b) = ($260.30^{\circ}$, $-3.16^{\circ}$)}.
\label{fig:maps}
\end{center}
\end{figure*}

In the study of the morphology of an extended source, a major requirement 
is to have the best possible angular resolution. Therefore, we decided to restrict 
our LAT dataset to events with energies above 800~MeV, which further reduces the Galactic diffuse background. 
Figure~\ref{fig:maps} presents the LAT Test Statistic map of off-pulse
emission in the Vela region. The Test Statistic (TS) is defined as twice 
the difference between the log-likelihood $L_1$ obtained by fitting 
a source model plus the background model to the data, and the log-likelihood 
$L_0$ obtained by fitting the background model only, i.e TS = 2($L_1$ - $L_0$). 
This skymap contains the TS value for a point source at each map location,
thus giving a measure of the statistical significance for the detection of 
a gamma-ray source in excess of the background. Note that the
pulsar (cross) is quite faint in this phase interval. 
The skymap shows bright emission south of PSR B0833$-$45 with a 
fainter extension to the east. This gamma-ray complex lies within Vela-X;
in particular it is contained within the region that remains strong at
high radio frequencies (denoted by the WMAP flux contours, see discussion).
An additional source, still unidentified but coincident with the North-Eastern part of 
the supernova remnant Puppis~A, is also visible at position (l, b) = ($260.30^{\circ}$, $-3.16^{\circ}$) 
with a TS value of 34.6. This source is taken into account in the spectral analysis.

\begin{figure*}[ht!!]
\begin{center}
\includegraphics[angle=0,scale=.7]{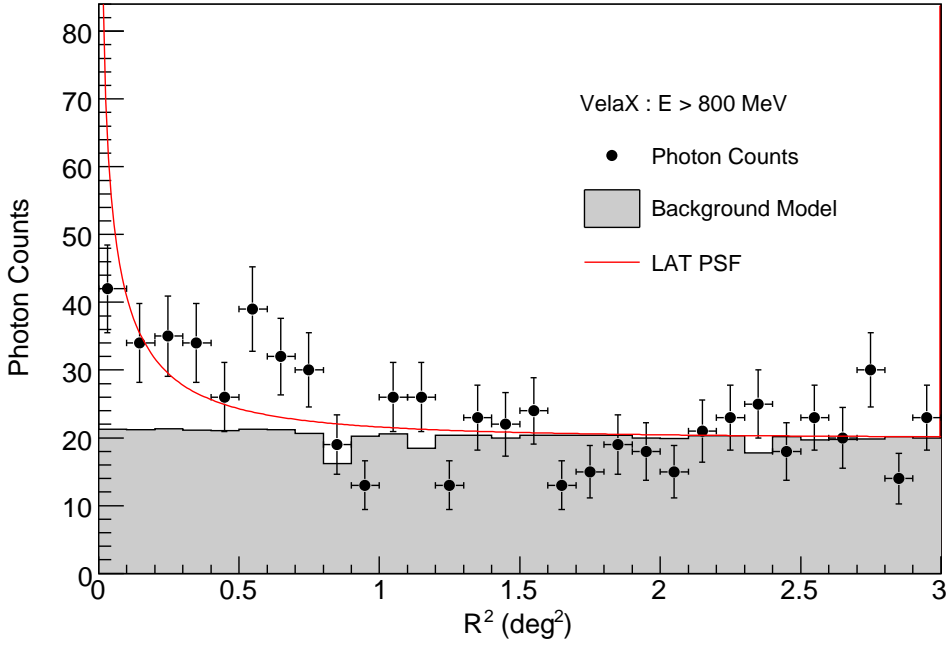} 
\caption{Radial profile of the LAT data about the best fit position 
provided by Sourcelike for a point source (l, b) = ($263.03^{\circ}$, $-3.27^{\circ}$) 
as reported in Table~\ref{tab:sourcelike} (E$>$800~MeV). The LAT PSF is 
overlayed as a red solid line for comparison. The background model is presented 
as a grey histogram and the black dots represent the LAT data.}
\label{fig:radprof_velax}
\end{center}
\end{figure*}

We determined the source extension using Sourcelike with a uniform
disk hypothesis (compared to the point-source hypothesis). The results
of the extension fits are summarized in
Table~\ref{tab:sourcelike}. The difference in TS between the uniform
disk and the point-source hypothesis is 47.9 (which converts into a
significance of $\sim 7 \sigma$ for the source extension) for 800~MeV $<$ $E$ $<$ 20~GeV, 
which demonstrates that the source is significantly extended with respect to the 
LAT point spread function (PSF). The fit extension has a radius of $0.88^{\circ} \pm 0.12^{\circ}$. 
We support this conclusion in Figure~\ref{fig:radprof_velax}, showing the 
radial profile for the LAT data above 800~MeV (from the best source location
determined for a point source fit) and comparing this with the 
LAT PSF.

We have also examined the correspondence of the gamma-ray emission
with different source morphologies by using gtlike with assumed
multi-frequency templates. For this exercise we compared the TS of the
point source and uniform model parameters provided by Sourcelike with
values derived when using morphological templates from the H.E.S.S.\
gamma-ray excess map~\citep{HESSVelaX} and the WMAP radio images at 
61-GHz (archival data, see section~\ref{mw}). The resulting
Test Statistic values obtained from our maximum likelihood fitting are
summarized in Table~\ref{tab:ts}. Fitting a uniform disk to the data using the
best location and size provided by Sourcelike improves the TS by 40.4 in
comparison to the point-source hypothesis, comparable to the
improvement in TS between $D$ and $PS$ models in
Table~\ref{tab:sourcelike}. Replacing the disk with spatial template
provided by the H.E.S.S. observations decreases the TS with respect to
the disk hypothesis ($\Delta$TS = $-$31.3), implying that the LAT
emission does not correspond well to the TeV flux. In contrast, using
the radio contours as spatial template improves the value of the Test
Statistic, but only by $\Delta$TS = +11.7. Thus while the best match
is with the radio morphology, as expected from the double electron
population scenario, we cannot (at high significance) rule out a
simple disk morphology.

\placetable{1}
\begin{table*}[ht]
\centering
\begin{tabular}{|c|c|c|c|c|c|c|c|}
\hline\hline 
Model & Name & Energy band (GeV) & l($^\circ$) & b ($^\circ$) & Radius ($^\circ$) & $\Delta$TS\\
\hline
Point Source & $PS$ & 0.8 - 20.0 & 263.03 & -3.27 & & \\
Disk & $D$ & 0.8 - 20.0 & 263.34 & -3.11 & 0.88 $\pm$ 0.12 & 47.9 \\
\hline
\end{tabular}
\caption{\label{tab:sourcelike} Centroid and extension fits to the LAT data for Vela-X using Sourcelike for events with energies above 800~MeV. }
\end{table*}

\begin{table*}[ht]
\begin{center}
\begin{tabular}{|c|c|c|}
\hline\hline
Model & Name & TS \\
\hline
Point Source & $PS$ & 44.0\\
Disk & $D$ & 84.4\\
HESS & & 53.1\\
WMAP 61 GHz & & 94.0\\
\hline
\end{tabular}
\caption{Comparison of model likelihood fitting results with gtlike for 
events with energies above 800~MeV. For each model, we give the name, 
and the Test Statistic value (TS).}\label{tab:ts}
\end{center}
\end{table*}

\subsection{Spectral analysis}
\label{spectrum}
\begin{figure*}[ht!!]
\begin{center}
\includegraphics[angle=0,scale=.7]{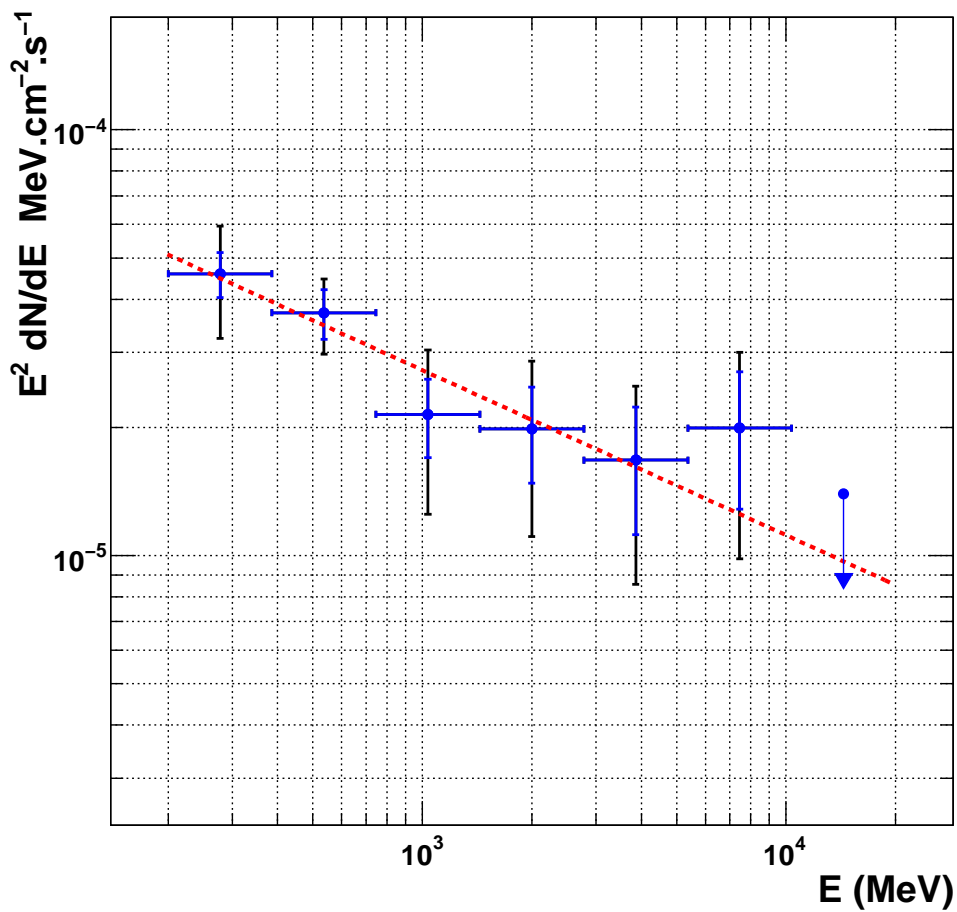} 
\caption{Spectral energy distribution of Vela-X renormalized to the total phase interval. 
The LAT spectral points are obtained using the maximum likelihood method described in section~\ref{spectrum} 
into 7 logarithmically-spaced energy bins. The statistical errors are shown in blue, 
while the black lines take into account both the statistical and systematic errors as discussed in section~\ref{spectrum}. 
The red dotted line presents the result obtained by fitting a power-law to the data in 
the 200~MeV-20~GeV energy range using a maximum likelihood fit. A 95~\% C.L. upper limit is computed 
when the statistical significance is lower than 3~$\sigma$.}
\label{fig:spec_velax}
\end{center}
\end{figure*}

The \emph{Fermi}-LAT spectral points were obtained by dividing the 200 MeV -- 20 GeV 
range into 7 logarithmically-spaced energy bins and performing a maximum likelihood spectral 
analysis in each interval, assuming a power-law shape for the source. For this analysis we used 
the uniform disk model from Table~\ref{tab:sourcelike} to represent the gamma-ray emission 
observed by the LAT, as discussed in section~\ref{morpho}. Assuming this spatial shape, the gamma-ray 
source observed by the LAT is detected with a significance of 14$\sigma$ in the 200 MeV -- 20 GeV range. 
The result, renormalized to the total phase interval, is presented in Figure~\ref{fig:spec_velax}. 
To determine the integrated gamma-ray flux we fit a power-law spectral model to the data in the energy 
range 200~MeV -- 20~GeV with a maximum likelihood analysis. This analysis is more reliable 
than a direct fit to the spectral points of Figure~\ref{fig:spec_velax} since it accounts 
for Poisson statistics of the data. The spectrum of Vela-X between 200~MeV and 20~GeV, assuming the uniform disk 
model from Table~\ref{tab:sourcelike}, is well described by a power-law with a spectral index 
of 2.41 $\pm$ 0.09 $\pm$ 0.15 and an integral flux above 100~MeV of (4.73 $\pm$ 0.63 $\pm$ 1.32)$\times 10^{-7}$ cm$^{-2}$~s$^{-1}$ 
(renormalized to the full phase interval). This is in agreement with results obtained 
independently using Sourcelike. The first error is statistical, while the second represents 
our estimate of systematic effects as discussed below. No indication of a spectral cut-off at 
high energy can be detected with the current statistics. This result takes into 
account the gamma-ray emission from the source coincident with Puppis~A, which
was well modeled as a point source emitting a power-law of spectral index 
1.97 $\pm$ 0.16 and integral flux above 100~MeV of (0.43 $\pm$ 0.16)$\times 10^{-7}$ 
cm$^{-2}$~s$^{-1}$ (statistical errors only).

As an attempt to estimate the level of pulsed emission in the off-pulse window, we fitted a 
point source at the position of the Vela pulsar, in addition to the uniform disk representing Vela-X. 
We derived an integral flux above 100~MeV of $\sim 3 \times 10^{-8}$ cm$^{-2}$~s$^{-1}$ for the point source, 
which represents $\sim 6$\% of the flux of Vela-X.

Fitting a point source only at the position of the Vela pulsar, we get a spectrum well 
described by a power-law with a spectral index of $2.98 \pm 0.16$ and an 
integral flux above 100~MeV of $(1.48 \pm 0.25) \times 10^{-7}$ cm$^{-2}$~s$^{-1}$ 
for the phase interval 0.7 - 1.0 (statistical errors only). This low flux is in agreement with the upper 
limit reported in~\cite{VelaPulsarLAT}.

Three different systematic uncertainties can affect the LAT flux estimation. The main systematic 
at low energy is due to the uncertainty in the Galactic diffuse emission since 
Vela-X is located only $2^{\circ}$ from the Galactic plane in a region of dense 
molecular clouds. Different versions of the Galactic diffuse emission generated by GALPROP 
were used to estimate this error. The difference with the best fit diffuse model is
found to be $\le 6$\%. By changing the normalization of the Galactic diffuse model artificially 
by $\pm 6$\%, we estimate this systematic error to be 25\% (0.2 - 0.4 GeV), 14\% (0.4 - 0.8 GeV) 
and $<$10\% ($>$ 0.8 GeV). The second systematic is related to the morphology of the LAT source. 
The fact that we do not know the true gamma-ray morphology introduces another 
source of error that becomes dominant when the size of the source is larger than the PSF, 
i.e above 600~MeV for the case of Vela-X. Different spatial shapes have been used to estimate this systematic
error: a disk, a Gaussian and the radio templates. Our estimate of this 
uncertainty is $\sim$25\% between 600 MeV and 1~GeV and 30\% above 1~GeV. 
The third uncertainty, common to every source 
analyzed with the LAT data, is due to the uncertainties in the effective area. 
This systematic is estimated by using modified instrument response functions 
(IRFs) whose effective area bracket that of our nominal IRF. These `biased' IRFs 
are defined by envelopes above and below the nominal dependence of the effective 
area with energy by linearly connecting differences of (10\%, 5\%, 20\%) at 
log(E) of (2, 2.75, 4) respectively.
We combine these various errors in quadrature to obtain our best estimate
of the total systematic error at each energy and propagate through to the fit model parameters.

\subsection{Supporting Multi-wavelength Measurements}
\label{mw}

	As a means of better understanding the Vela PWN, we compiled and analyzed multi-wavelength 
data corresponding to the longer wavelength synchrotron counterparts 
of the sub-GeV-peak (halo) and TeV-peak (cocoon) Compton emission. 
Although their morphologies do vary with waveband, we have attempted to
form the SEDs of the {halo and cocoon of Vela-X by using consistent
apertures. This is important in this complex region and has, apparently,
not been the practice in some previous studies. Vela-X itself has
been traditionally studied at low radio frequencies where the spatial resolution
is very poor. However, in an 8.4 GHz Parkes image (Fig 2 of~\cite{hales}), a 
$\sim 2.5^\circ \times 1.5^\circ$ region of bright filamentary emission is 
visible, roughly coincident with the extended LAT flux. 
We examined archival 5-year WMAP sky maps\footnote {http://lambda.gsfc.nasa.gov} and find that 
this region appears as a distinct concentration in the
WMAP all-sky images at 23-, 33-, 41-, 61-, and 94-GHz. As the
resolution increases to higher frequencies it is increasingly separated
into eastern and western sub-regions, both well south of the Vela pulsar. We measured a flux 
for each energy band and estimated a flux error (dominated by
the uncertainty in the background estimation) using the region defined in Figure~\ref{fig:regions}.  
This concentration is also clear in the 0.4 GHz all sky maps of~\cite{haslam}, which
provide a low-frequency point. The flux measurements are plotted in 
Figure~\ref{fig:sed_velax} .
We were not able to extract a reliable flux estimate from the 8.4 GHz
map. We estimate the flux density spectral index for this region of
Vela-X as $\alpha = 0.5\pm0.05$, similar to but steeper than the 
$\alpha=0.39\pm0.03$ index measured over 0.03-8GHz for a much larger 
region covering all of Vela-X~\citep{alvarez}. 
The component measured here is $\sim 5 \times $ fainter. 
Additional mm and IR measurements would be very helpful in extending
the spectrum and searching for the expected synchrotron peak at
$\sim$mm wavelengths.

In the X-ray band many authors have estimated the spectrum of the 
cocoon region, starting with the {\it ROSAT} analysis of
\citet{mo95}. More detailed fitting with {\it ASCA} \citep{mo97,
  hornsetal06} showed that the emission must consist of an optically thin thermal plasma 
(typically a mekal thermal plasma model) with $kT \approx 0.3$\,keV 
plus a power law ($\Gamma \approx 2.0\pm 0.3$) component.  More
recently, \cite{lamassa} have analyzed XMM data of the bright central
portion of the cocoon, and fit a thermal plasma
($kT=0.48_{-0.06}^{+0.05}$ keV) plus power law ($\Gamma =
2.3\pm0.04$). The 0.2--6.5 keV power-law flux that they find
corresponds to $9.5 \pm 1.2 \times 10^{-11} \rm \, erg \, cm^{-2} \,
s^{-1}$), when scaled up to the area of the bright H.E.S.S. emission
considered here.  These authors also fit a hydrogen column density of
$n_H = 1.6_{-0.2}^{+0.3} \times 10^{20} \, \rm cm^{-2}$, which we
shall adopt for our analysis. All of the analyses extending above
2\,keV have been forced to measure only portions of the long $\sim
1.5^\circ$ cocoon structure. 
A number of older X-ray (Einstein \citep{1985ApJ...299..828H};
HEAOA-4 \citep{1984ApJS...54..581L})
and soft gamma-ray 
(OSSE \citep{dejager96};  
BeppoSAX \citep{mangano})
observations of the Vela
plerion possess a large enough field of view to encompass the majority of Vela-X.  
Yet the spectral extraction regions of these observations are centered on the Vela pulsar, such that
the bright inner PWN contaminates the low surface brightness
extended nebula and hardens the net spectrum.  We therefore
refrain from using such archival data as an estimate of the Vela-X spectrum.  

We  made a first attempt to improve such measurements by fitting
to the combined emission in several {\it ASCA} GIS2/3 pointings that cover the
bulk of the cocoon, as presented in Figure~\ref{fig:regions}. 
Data set 25038000 (76 ks livetime) covered the northern region
while data sets 23043000 and 23043010 (combined livetime 134 ks)
cover the southern region. Using XSelect version 2.4, we extract data sets
from two $20 \arcmin \times 20 \arcmin$ regions, one each in the north and south
which largely covered this region. The large GIS FOV allowed us to select background
regions well outside of the cocoon but on the same detector. We assumed a fixed absorption
$n_H = 1.6 \times 10^{20} \, \rm cm^{-2}$ and fit a mekal thermal plasma
plus power law to the combined data. The thermal component is fit with
$kT = 0.51_{-0.04}^{+0.05}$ keV (single parameter 90\% errors) over all datasets;
no significant variation is seen in $kT$ for independent fits to the
northern and southern regions. To best constrain the power-law component
we restricted the fit to the 2-10\,keV range -- here separate fits gave
$\Gamma=1.97_{-0.05}^{+0.06}$ in the north and a slightly softer $2.15\pm0.10$
in the south, providing weak evidence for aging of the electron population 
as one moves along the cocoon.
Finally for comparison with the H.E.S.S. emission,
we fit to the combined regions, obtaining an average index of $2.06 \pm 0.05$ and
2--10 keV flux of $6.7 \pm 0.4 \times 10^{-11} \rm \, erg \, cm^{-2} \, s^{-1}$
(scaling up the flux in our extraction aperture to the area of the bright H.E.S.S.
emission).
This corresponds to a 0.3--7 keV flux of $1.4\pm0.1\times 10^{-10}{\rm \, erg 
\, cm^{-2} \, s^{-1}}$, in good agreement with previous estimates. The spectral
energy distribution (SED) points from the 2--10 keV power-law portion of this fit are
plotted in Figure~\ref{fig:sed_velax}.

	Finally, we wish to check for X-ray emission from the larger halo portion
of Vela-X covered by the radio/LAT component. This very large region is presently
well covered only by the ROSAT All-Sky Survey (RASS), which is strongly
dominated by the bright thermal emission of the Vela SNR, particularly at low energy.
To produce a bound on the flux we measured the counts within the radio/LAT region
in the hard-band 0.5--2.0 keV RASS image, subtracting background from appropriate
surrounding regions. No significant excess counts were found and we convert the
upper bound on the flux of a $\Gamma=2$ power-law component using WebPIMMS, obtaining
$2.5 \times 10^{-11} \rm \, erg \, cm^{-2} \, s^{-1}$. This bound is shown by an
arrow in Figure~\ref{fig:sed_velax}.

\section{Discussion}
\label{discussion}

Different scenarios have been proposed to interpret the multi-wavelength observations
of Vela-X. Horns et al. (2006) proposed a hadronic model wherein the gamma-ray 
emission is the result of the decay of neutral pions produced in proton-proton 
collisions in the cocoon. However, this model requires a particle density larger than 
$0.6$~cm$^{-3}$, which seems disfavored by the recent best fit estimate of thermal particle density of
$\sim 0.1$~cm$^{-3}$ using XMM observations~\citep{lamassa}. 
\citet{lamassa} proposed a leptonic model with radio and X-ray emissions 
resulting from synchrotron radiation and gamma-ray emission arising from inverse Compton scattering. 
In this model, the authors need a 3-component broken power-law 
to describe the electron population and adequately fit the data. A model with a 
single break can also reproduce the multi-wavelength data if a 
separate electron population produces the radio emission~\citep{dejager}. 
In this case, the morphology of the gamma-ray emission observed by {\it Fermi} 
should be similar to that in the radio since they are produced by the 
same electron population. In the model of \citet{dejager}, the low energy
electron component has a total energy of $4\times 10^{48}$ erg, while the
X-ray/TeV-peak component has a total lepton energy of $2\times10^{46}$ erg. Both
employ a magnetic field of $5 \, \mu$G.   

	Our new \emph{Fermi}-LAT spectrum and the improved flux estimates for
the radio and X-ray emission from the two components of our SED 
(Figure~\ref{fig:sed_velax}) allow considerable progress in constraining the model
parameters. 
First, the steep LAT spectrum disfavors the hadronic scenario. While
the VHE gamma-ray data can be adequately fit with gammas from pion
decay, neither the ASCA nor the LAT data can be accounted for by
secondary electrons. We therefore require a three-component injection
(one hadron and two lepton) in this case, along with a quite high magnetic field in
the cocoon in order to suppress IC scattering of X-ray emitting
electrons from providing the dominant source of VHE gamma-rays.
As noted by \citet{dejager}, the SED strongly supports a two-component leptonic
model. We have computed the SEDs from evolving power-law electron populations, one each for the 
X-ray/VHE-peak cocoon and radio/sub-GeV-peak halo. In both regions an exponentially 
cut-off power law is injected at constant luminosity and evolved for the 11\,kyr estimated 
lifetime of the Vela pulsar, subject to synchrotron and Klein-Nishina adjusted Compton losses. 
We ignore any possible adiabatic losses to the electron population, since these are quite
uncertain and may, in any case, be offset by the compression from the SNR reverse shock.
IC seed fields include CMB, far IR (temperature 25 K, density 0.4 eV$\rm \, cm^{-3}$) 
and starlight (temperature 6500 K, density 0.4 eV$\rm \, cm^{-3}$ \citep{dejager}), 
reasonable for the the locale of Vela-X \citep{pms06}. For each region we vary the 
magnetic field, power-law cutoff energy, power-law index, and total lepton energy; 
we find the best model fit by minimizing the weighted chi-squared statistic between 
model and data points.  For each parameter 90\% one-dimensional errors are subsequently 
calculated by varying the best-fit value of the given parameter until chi-squared increases by 2.71.
	The $\alpha=0.5$ halo radio spectral index suggests an electron 
power-law index close to the classical $p = 2\alpha +1 = 2$. The synchrotron/Compton
peak ratio of the cocoon implies a $B=4 \,\mu$G field, with small uncertainty. In fact we 
adequately match the SED of both components with this field and an
$E^{-2}$ spectrum. However for the cocoon region we require a 
600 TeV exponential cut-off and total energy $1.5 \times 10^{46}$ erg, while the 
halo requires a lower 100\,GeV exponential cutoff and a total energy
of $5\times10^{48}$ erg. The peaks of the cocoon component are controlled by
the cooling break. The halo population does not cool appreciably 
during the pulsar lifetime and the peak energies are controlled
by the exponential cut-off of the injected spectrum. The X-ray upper limit on this 
component is not constraining. Note that we do not require a mid-range break in the
injected spectrum for either component. 

	With so many free parameters, such SED fits are usually illustrative, rather
than constraining. However, with our new LAT detection and improved low energy measurements we
are testing the plausible injection spectrum for the Vela-X PWN. We list the parameters determined 
by chi-squared fits to the multi-wavelength data and single-parameter fit errors in Table~\ref{fits}. 
The cocoon emission evidently represents significantly cooled electrons, dominated by relatively
recent injection of high energy electrons from the pulsar and its termination shock. 
The halo component, on the other hand, represents old electrons -- these are easily produced over the lifetime
of the pulsar for any initial spin period $\le 60$\,ms. Although it would be very
interesting to push the LAT spectral measurement to lower energy, where
the halo spectrum may peak, this will prove very difficult even with more exposure,
given the poor low energy PSF. On the other hand, extension of the radio spectrum
through the mm band promises to constrain the high energy cut-off of the halo 
electron spectrum. For the cocoon component, scheduled {\it XMM} mapping of this
region should provide appreciable improvement in the spectral measurements of the 
non-thermal X-rays and may
extend to low enough energy to probe the synchrotron peak. With such refined constraints
we should have a quite detailed knowledge of the bulk injection from the pulsar
and its termination shock. In turn, it may be hoped that this, and similar
measurements of other PWNe, will help us understand the physics of these relativistic outflows.

\begin{table*} [h]
\begin{center}
\begin{tabular}{|c|c|c|c|c|c|}
\hline \hline
Component & $B$ ($\mu$G) & $E_c$ (eV) & $\Gamma$ & $E_{\rm tot}$ (erg) & $\chi^2$/DoF \\  \hline
& & &  & & \\
Halo  & 3.93$^{+0.46}_{-0.38}$ & 1.01$^{+0.07}_{-0.13}\times 10^{11}$ & 1.97$^{+0.02}_{-0.02}$ & 
      $5.05^{+0.45}_{-0.56} \times 10^{48}$ & 10.7/9 \\ 
 & & & & &\\
Cocoon& 3.80$^{+0.10}_{-0.08}$ & 5.69$^{+0.16}_{-0.33}\times 10^{14}$ & 1.998$^{+0.003}_{-0.001}$ & 
      $1.50^{+0.01}_{-0.05} \times 10^{46}$ & 57.7/15 \\ 
& & & & & \\
                       \hline
 \end{tabular}
 \end{center}
 \caption{\label{fits} Multiwavelength SED fit to the PWN components as seen in Figure~\ref{fig:sed_velax}.}
\end{table*}

\begin{figure*}[ht!!] 
\begin{center}
\includegraphics[angle=0,scale=.7]{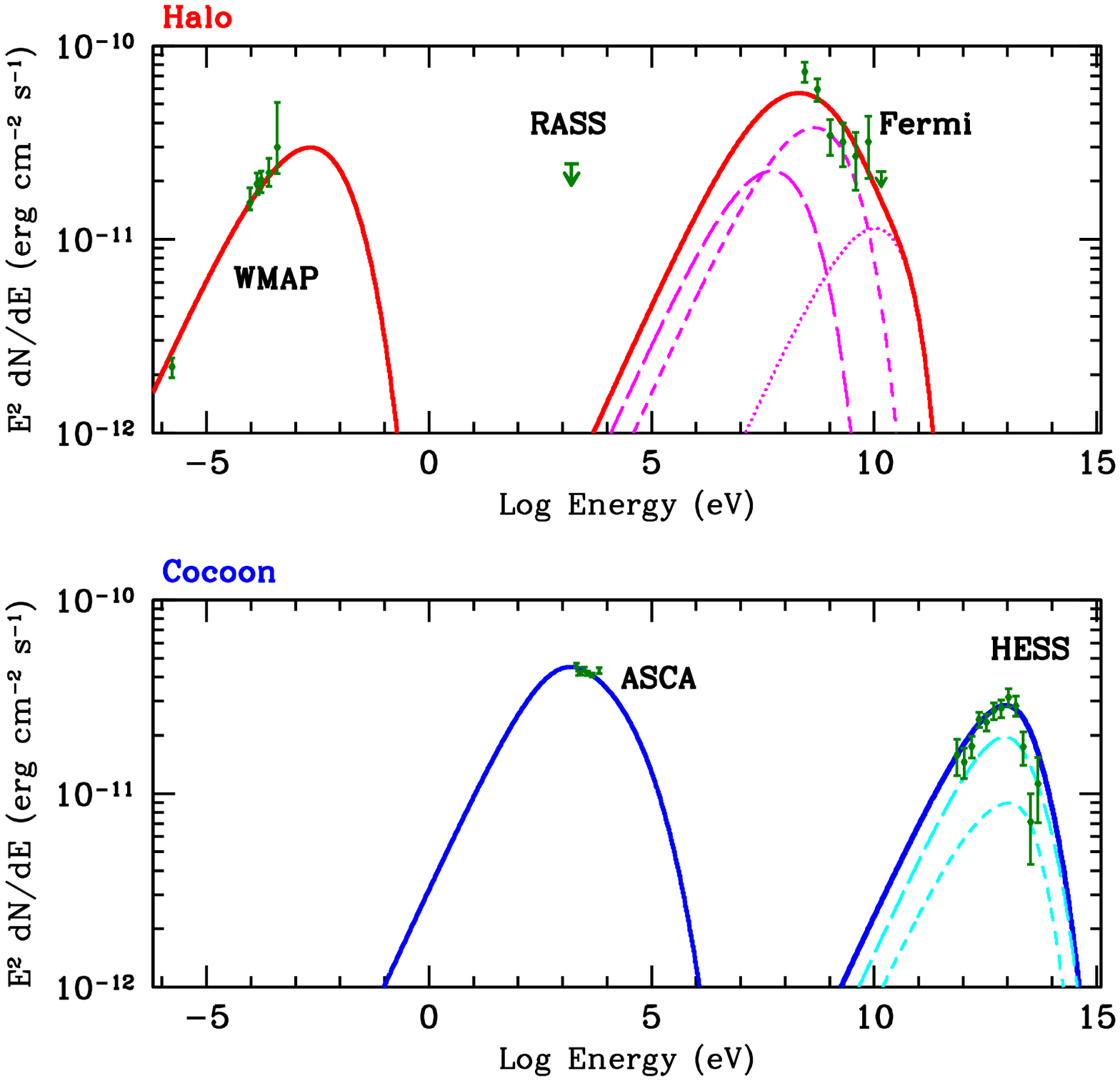} 
\caption{Spectral energy distribution of regions within Vela-X from 
radio to very high energy gamma-rays. 
{\bf Upper panel:} Emission from the low energy electron population (halo). WMAP and GeV gamma-ray points (this paper) are
for the large radio-bright portion of Vela-X. The ROSAT upper limit (this paper) 
on the soft X-ray flux of this region is also shown by an arrow. 
The Compton components from scattering on the CMB (magenta long dashed line), dust emission 
(magenta dashed line) and starlight (magenta dotted line) are shown. 
{\bf Lower panel:} Synchrotron and Compton emission 
from the high energy electron population (cocoon). X-ray (ASCA observations, this paper) and
very high energy gamma-ray (Aharonian et al. 2006) points are also from the cocoon region. 
Only CMB (cyan long dashed line) and dust (cyan dashed line) 
scattered flux is shown as the starlight is Klein-Nishina suppressed. 
}
\label{fig:sed_velax}
\end{center}
\end{figure*}


\acknowledgments
The \emph{Fermi} LAT Collaboration acknowledges generous ongoing
support from a number of agencies and institutes that have
supported both the development and the operation of the LAT as well as 
scientific data analysis. These include the
National Aeronautics and Space Administration and the Department
of Energy in the United States, the Commissariat \`a
l'Energie Atomique and the Centre National de la Recherche
Scientifique / Institut National de Physique Nucl\'eaire et de
Physique des Particules in France, the Agenzia Spaziale Italiana,
the Istituto Nazionale di Fisica Nucleare, and the Istituto
Nazionale di Astrofisica in Italy, the Ministry of Education,
Culture, Sports, Science and Technology (MEXT), High
Energy Accelerator Research Organization (KEK) and Japan
Aerospace Exploration Agency (JAXA) in Japan, and the K.
A. Wallenberg Foundation and the Swedish National Space
Board in Sweden.
Additional support for science analysis during the operations 
phase from the following agencies is also gratefully acknowledged: 
the Instituto Nazionale di Astrofisica in Italy and the Centre National d'\'Etudes Spatiales in France.


\end{document}